      [1999/12/01 v1.4c Il Nuovo Cimento]
\documentclass[varenna]{cimento}

\newcommand{\be}{\begin{equation}}

\newcommand{\ee}{\end{equation}}
\newcommand{\bea}{\begin{eqnarray}}
\newcommand{\eea}{\end{eqnarray}}

\newcommand{\nn}{\nonumber}
\newcommand{\de}{\partial}

\def \nn{\nonumber}

\def\slr#1{\setbox0=\hbox{$#1$}           
   \dimen0=\wd0                                 
   \setbox1=\hbox{/} \dimen1=\wd1               
   \ifdim\dimen0>\dimen1                        
      \rlap{\hbox to \dimen0{\hfil/\hfil}}      
      #1                                        
   \else                                        
      \rlap{\hbox to \dimen1{\hfil$#1$\hfil}}   
      /                                         
   \fi}

\def\be{\begin{eqnarray}}
\def\ee{\end{eqnarray}}

\def\eeas{\end{eqnarray*}}
\def\bea{\begin{eqnarray}}
\def\eea{\end{eqnarray}}
\def\slash#1{#1\!\!\!/}
\title{Introduction to Color Superconductivity}
\author{G. Nardulli}

\institute{Dipartimento di Fisica, Universit\`a di Bari, I-70126
Bari, Italia\\ I.N.F.N., Sezione di Bari, I-70126 Bari, Italia}
\begin{document}

\maketitle

\begin{abstract}At high nuclear density  and small temperature
due to the asymptotic freedom property of Quantum ChromoDynamics
and to the existence of an attractive channel in the color
interaction, diquark condensates might be formed. Since these
condensates break the color gauge symmetry, this phenomenon has
the name of color superconductivity. In the last few years this
has become a very active field of research. While a direct
experimental test is still missing, color superconductivity might
have implications in  astrophysics because  for some compact
stars, e.g. pulsars, the baryon densities necessary for color
superconductivity can probably be reached.
\end{abstract}

\section{Nuclear matter and QCD}

At small nuclear density and low temperature the relevant degrees
of freedom of nuclear matter are the hadrons: nucleons, hyperons,
mesons. At high density and/or high temperature the fundamental
fields are those of quarks, spin 1/2 particles carrying electric
charge and a quantum number called color,
 and those of gluons, massless spin 1 particles carrying color in eight
 possible different ways. The corresponding theory is
 called Quantum ChromoDynamics
(QCD). It resembles Quantum Electrodynamics (QED), where the role
of the fundamental fermion is played by the electron and that of
the eight  gluons by the  photon (carrying no charge and therefore
unique). Similarly to QED, QCD is a gauge theory based on a local
unitary gauge group. For QCD it is $SU(3)$, while QED is based on
the abelian group $U(1)$. Quarks belong to the fundamental and
gluons to the adjoint representation of $SU(3)$, which is why each
quark has three  and each gluon eight possible colors. Besides
color, quarks carry another quantum number, flavor, that affects
their mass. Six different flavors are known, but we will consider
here only the lightest quarks, called $up$, $down$ and $strange$
(with $m_u\simeq m_d\ll m_s$). A peculiar property of QCD is
Asymptotic Freedom (AF), which means that at large momenta the
effective QCD coupling constant becomes smaller and smaller with
increasing momenta.

In the ordinary world, i.e. at zero temperature and zero density,
the study of the QCD degrees of freedom is made difficult  by the
fact that quarks and gluons are $confined$ inside hadrons.
Studying the QCD vacuum state under extreme conditions can
therefore help our understanding of strong interactions, because
their description is simpler in these regimes, due to the AF
property of QCD. In particular one might perform the study in two
very different regimes: high temperature and small density or
small temperature and high density. The first regime is presently
studied at the Relativistic Heavy Ion Collider at Brookhaven, USA,
and will be investigated in the near future at the Large Hadron
Collider at CERN. The other regime is not accessible for the
present at the accelerators, but can be indirectly investigated by
the study of the properties of compact stars such as pulsars,
where both conditions, small temperatures and high densities, are
likely to occur.

As we noted  quarks are spin 1/2 fermions with  three colors,
conventionally called red, green, blue $(r,g,b)$ and, if light, in
three flavors ($u,d,s$). Since they belong to the representation
$\bf 3$ of the color symmetry group, when two quarks scatter  they
are either in a sextet (color symmetric) or in antitriplet (color
antisymmetric) state. QCD interactions are mediated by one gluon
exchange (in the perturbative regime). In the antitriplet channel
this interaction is attractive. As a consequence of the Cooper's
theorem the perturbative vacuum is unstable and one expects the
formation of Cooper pairs of quarks (diquarks) and color
superconductivity (this is so  because the condensate  is colored,
not white as in the quark-antiquark channel, relevant at zero
density). Moreover the two quarks scatter preferably in the spin 0
channel, since this corresponds to a larger phase space. At
extreme densities, considering QCD with 3 flavors, the attractive
interaction in color antisymmetric channel, together with $S=0$
implies, by the Pauli principle, antisymmetry in flavor. This is
the basic idea of the models that we discuss in the next section.

\section{The true vacuum}Since the
naive vacuum is unstable we should try to find  the true
superconducting vacuum state.  To do that we have to show that the
formation of a quark-quark condensate occurs because of the color
interaction.
 The interaction term is fundamentally provided by gluon exchanges and to compute it and derive the gap equation
 one can follow different approaches, such as, for example,
 the one of Bailin and Love in \cite{others} based on
 the resummation of bubble diagrams (for previous work
 see references in \cite{others}).
 For this method and similar approaches we refer to the reviews
\cite{rassegne}.

For pedagogical purposes it is however sufficient to consider a
Nambu-Jona Lasinio \cite{NJL} (NJL) interaction  (4 fermion
interaction) that mimics the QCD interaction. In this
approximation the hamiltonian is: \be H=\int d^3x\left(\bar
\psi(i\,\slash\partial+\mu\gamma_0)\psi+G\bar \psi\gamma^\mu
T^A\psi\bar \psi\gamma_\mu T^A\psi\right)\ . \label{true}\ee Here
$G$ is a coupling constant with dimensions (mass)$^{-2}$ and $T^A$
 are $SU(3)$ generators.
 Notice that a form factor is included in the definition, to take
into account the asymptotic freedom property of QCD. It would act
as an ultraviolet cutoff.

In order to obtain the gap equation we make a  mean field ansatz:
\be \psi^T_{\alpha i} C\psi_{\beta j}~\to \langle\psi^T_{\alpha i}
C\psi_{\beta j}\rangle . \label{condensatetrue0}\ee $C$ is the
charge conjugation matrix. Here and below Greek indices
$\alpha,\,\beta$ are color indices, Latin indices $i,\,j$ are
flavor indices. Let us now show that one expects for the vacuum
expectation value of the quark bilinear the result\be
\langle\psi^T_{\alpha i} C\psi_{\beta j}\rangle=
\epsilon_{\alpha\beta I}\epsilon_{ijK}\frac{\Delta_{IK}}{2G}\
\label{condensatetrue}\ee (sum over repeated indices). It is
useful to note that, by evaluating (\ref{true}) in mean field
approximation, i.e. substituting to a pair of fermion fields in
(\ref{true}) their mean value given by (\ref{condensatetrue}) and
(\ref{condensatetrue0}), the interaction term takes the form
\be\sim\Delta\psi\psi\ +\ h.c.\ ,\ee i.e. the fermions acquire a
Majorana mass.

To prove (\ref{condensatetrue}) we note that,
 as we have already discussed, QCD interaction favors antisymmetry in color; the pair has  zero
angular momentum (in this way the entire Fermi surface is
available and the effect becomes macroscopic); if it has also spin
zero the fermion pair must be in an antisymmetric state of flavor
to produce an antisymmetric wavefunction. Neglecting the sextet
contribution, which is small, leaves us with
(\ref{condensatetrue}).

Clearly the form  of the matrix
 $\Delta_{IK}$ corresponding to the true vacuum state depends on dynamical
 effects. In the literature various cases have been
 discussed:
\begin{enumerate}
 \item  2 flavor SuperConductivity (2SC): $\Delta_{IK}
 =
 \Delta\delta_{I3}\delta_{K3}$ ~~~~~\cite{alford}~;
 \item Color-Flavor-Locking (CFL): $\Delta_{IK}
 =\delta_{IK}\Delta$ ~~~~~\cite{wilczekcfl}~;
\item  gapless Color-Flavor-Locking (gCFL): $\Delta_{IK}
 =\delta_{IK}\Delta_K$
 ~~~~~\cite{Alford:2003fq,Casalbuoni:2004tb}~;
 \item   2 flavor LOFF phase: $\Delta_{IK}
 =
 ~\delta_{I3}\delta_{K3}\Delta({\bf r})$~~~~~\cite{LOFF2,Alford:2000ze,Casalbuoni:2003wh,Casalbuoni:2004wm}
~;
 \item  3 flavor LOFF phase:$\Delta_{IK}
 =\delta_{IK}\Delta_K({\bf r})$~~~~~\cite{Casalbuoni:2005zp}~.\end{enumerate}
It is an incomplete list, for other results see
\cite{Rapp},\cite{shovkovy} and references in \cite{rassegne}.

In the case of the LOFF phases the $r$ dependence of
 the condensates produce an inhomogeneous type of
 superconductivity. Various  $\bf r$
 dependences have been discussed, e.g. the single plane wave or the
 sum of several plane waves (see below). In the cases of the gCFL phase and the 3 flavor LOFF phase
 one has to take into account the differences of quark masses
 (as we noted above, $m_s$ is significantly larger that the up and down masses)
 and the differences of quark chemical potentials
  ($\mu_u,\mu_d,\mu_s$ are in general different).
  This introduces more parameters in
  the description, but their number is reduced by imposing that the quark matter is electrically and color neutral
  (such constraint is automatically implemented in the CFL model).
  We will concentrate our attention on the above mentioned  models.
\section{Homogeneous color superconductivity}
To show that color superconductivity can exist, let us consider the
first two cases (2SC and CFL). The 2SC model corresponds to the
presence in the condensate of only the $u$ and $d$ flavors: this
means that this ansatz may be relevant in the situation
characterized by an intermediate chemical potential
$m_u,\,m_d\ll\,\mu\,\ll\, m_s$. Therefore one can imagine of having
taken the limit $m_s\to\infty$ and the theory is effectively a
two-flavor theory. The CFL model holds in the other limit
$m_u,\,m_d,~m_s\ll\,\mu$. It takes its name from the property\be
\sum_{K=1}^3\,\epsilon_{\alpha\beta
K}\epsilon_{ijK}\,=\,\delta_{\alpha i}\delta_{\beta
j}-\delta_{\alpha j}\delta_{\beta i}\ ,\label{locking}\ee by which
the color and flavor indices  are locked.

Let us now introduce $a,a^\dag,b,...$ annihilation/creation
operators of particles and holes:
  \be \psi_{\alpha
 i}=\sum_k u_k\left(a_{k}e^{ikx}+
 b^\dag_{k}e^{-ikx}\right)_{\alpha
i }\ .\ee

 The existence of the  condensate gives rise
 to the following hamiltonian
\bea H&=&\sum_{\vec k}\left(
|k-\mu|a^\dag_ka_k+(k+\mu)b^\dag_kb_k\right)\cr &+&\sum_{\vec
k}\frac{\Delta }2 e^{-i\Phi}a_k a_{-k}+\frac{\Delta}2
e^{+i\Phi}b^\dag_k b^\dag_{-k}+hc\eea where we have omitted for
simplicity any color-flavor indices as we wish to stress the
mechanism that produces the condensate. Let us now perform  a
unitary Bogoliubov  transformation. To do that  we introduce
annihilation and creation operators for quasiparticle and
quasiholes \bea y_k&=&\cos\theta a_k-e^{i\Phi}\sin\theta
a^\dag_{-k}\ ,\cr z_k&=&\cos\varphi b_k-e^{i\Phi}\sin\varphi
b^\dag_{-k}\ .\eea By an appropriate choice of the parameters of
the transformation we can transform the original hamiltonian in a
new hamiltonian describing a gas of non interacting quasiparticles
\be  H=\sum\left[ \epsilon_y(k) y^\dag_k y_k\,+\, \epsilon_z(k)
z^\dag_k z_k \right]\label{free}\ee  with \bea\label{dl}
\epsilon_y(k)&=&\sqrt{(k-\mu)^2+\Delta^2}\ ,\cr&&\cr
 \epsilon_z(k)&=&\sqrt{(k+\mu)^2+\Delta^2} \ .\eea
These equations show two effects:
\begin{enumerate}
\item The quasiparticles and quasiholes are free;
  \item in the dispersion laws of the quasiparticles a mass term
  proportional to $\Delta$ (the gap parameter) appears.
\end{enumerate}
 To obtain (\ref{free}), i.e. a free hamiltonian, the parameters have to be chosen as follows:
 \bea
\cos2\theta&=&\frac{|k-\mu|}{\sqrt{|k-\mu|^2+\Delta^2}} \ ,\cr
\cos2\varphi&=& \frac{k+\mu}{\sqrt{|(k+\mu)^2+\Delta^2}} \eea
 We note that, differently from the original $a_k,\,b_k$ operators
 that annihilate the false vacuum, the quasiparticles annihilation operators destroy the true vacuum
 \be y_k|0>=0,~~~~z_k|0>=0\ .\ee

   We have still to
prove that $\Delta\neq 0$. We do that and get an equation for
$\Delta $  by substituting  for $y_k, z_k$ in eq.
(\ref{condensatetrue}). We get in this way an integral equation that
has the following schematic form
 ({\it gap equation}):
 \be
  \Delta = C \int d^3k
  \Big[\frac{\Delta}{\sqrt{(k-\mu)^2+\Delta^2}}+
    \frac{\Delta}{\sqrt{(k+\mu)^2+\Delta^2}}\Big]\ .
\label{gap}\ee We see immediately that the origin of the instability
of the false vacuum lies in the first of the two terms in the r.h.s.
of (\ref{gap}): If $\Delta=0$ there is no compensation for the
divergence at $k=\mu$.

To be more quantitative let us consider the two models in more
detail. In the CFL model ($m_s= m_u= m_d=0$) all the $3\times 3=9$
quarks acquire a Majorana mass. The CFL condition gives two
different set of eigenvalues. The first one comprises 8 degenerate
masses $\Delta_1=\Delta_2=...=\Delta_8$ and the second set the non
degenerate mass $\Delta_0$. The actual values of the gaps depends
on the model and the approximations involved; typical values are
\be\Delta_1=...=\Delta_8=\Delta\approx\ 20 - 100\ {\rm MeV}\
,\hskip1cm \Delta_0\approx-2\Delta \ ,\label{gapcfl}\ee for
$\mu\approx 400 - 500$ MeV.

Let us observe that, since we start with a massless theory, the
QCD hamiltonian has, besides the local color symmetry $SU(3)$,
also a global chiral symmetry $SU(3)_L\times SU(3)_R$. The pairing
occurs for left-handed and right handed quarks separately.
Therefore these condensates break $SU(3)_c\otimes SU(3)_L\otimes
SU(3)_R$; finally, since the pair has baryonic number 2/3 (each
quark has baryonic number equal to 1/3 since the nucleon is made
up by three quarks), also $U(1)_B$ is broken. It should be
observed, however, that a diagonal $SU(3)_{c+L+R}$ symmetry
remains unbroken.

In the 2SC model ($m_s\gg\mu\gg m_u, m_d$) there are 4 massless
quarks (the up and down quarks with colors 1 and 2), while the
strange quark with any color
 and the $u$ and $d$ quarks with color 3 remain ungapped.
Numerical results for this case are qualitatively in agreement with
the results (\ref{gapcfl}) of the CFL model.

It is interesting to observe that the spontaneous breaking of
global symmetries implies the existence of Nambu Goldstone bosons;
for internal symmetries there are as many NGBs as there are broken
generators $G_a$. As they are massless, they are the lowest energy
quasiparticles of the effective theory. On the other hand, for
broken local gauge symmetries, by the Higgs-Anderson mechanism the
gauge bosons acquire masses and there are no NGBs. In the CFL
model the diquark condensate breaks  $SU(3)_L\times SU(3)_R\times
SU(3)_c\times U(1)_B$ to $SU(3)_{c+L+R}\times Z_2$. All the 9
quarks are massive and they belong to a $SU(3)_{c+L+R}$ singlet
and a $SU(3)_{c+L+R}$ octet. All the 8 gluons are massive and are
degenerate. There are 8+1 Nambu-Goldstone bosons. In the 2SC model
the condensate breaks $SU(3)_c\otimes SU(2)_L\otimes SU(2)_R\times
U(1)_B$  down to $ SU(2)_c\times SU(2)_L\times SU(2)_R$. While the
chiral group is unbroken, $SU(3)_c$ is broken to $SU(2)_c$.
Therefore 3 gluons remain massless and 5 acquire a mass. As to the
other quasiparticles, there is one would-be NGB associated to the
breaking of the axial symmetry; moreover of the 6 quarks (2
flavors in 3 colors) 4 are massive and 2 are massless.

One can show that, for arbitrarily large $\mu$, the CFL model is
favored \cite{Rajagopal:2000rs}; however, for intermediate $\mu$
other phases should be considered. At lower densities, in a range
presumably more relevant for the  study of compact stars, taking
into account strange quark mass effects implies that a new phase,
called gCFL (gapless CFL) phase, is favored. Let us therefore
consider now this new phase.

The Lagrangian for gluons and ungapped quarks with $m_u=m_d=0$ and
$m_s\neq0$ can be written as follows (color, flavor and spin
indices suppressed):
\begin{equation}
{\cal L}=\bar{\psi}\,\left(i\,D\!\!\! / -{\bf M}+ \bf{\mu}
\,\gamma_0\right)\,\psi \label{lagr1}
\end{equation}
where ${\bf M} = {\rm diag}(0,0,m_s) $ is the mass matrix  in flavor
space and the matrix of chemical potential is given by
 \be {\bf\mu}_{ij}^{\alpha\beta} = \left(\mu_b
\delta_{ij}- \mu_Q Q_{ij}\right)\delta^{\alpha\beta}+ \delta_{ij}
\left(\mu_3 T_3^{\alpha\beta}+\frac{2}{\sqrt 3}\mu_8
T_8^{\alpha\beta}\right) \ee ($i,j =1,3 $ flavor indices;
$\alpha,\beta =1,3 $ color indices). Moreover $T_3 = \frac 1 2 {\rm
diag}(1,-1,0)$, $T_8 = \frac{1}{2 \sqrt 3 }{\rm diag}(1,1,-2)$ in
color space and $Q= {\rm diag} (2/3,-1/3,-1/3)$ in flavor space;  $
\mu_Q$ is the electrostatic chemical potential; $\mu_3, \mu _8$ are
the color chemical potentials associated respectively to the color
charges $T_3$ and $T_8$; $\mu_b$ is quark chemical potential.

Working in the mean field approximation and neglecting the
antiquark (hole) contribution, one can derive the quasiparticle
dispersion law, whose knowledge allows the evaluation of the grand
potential. At zero temperature it is given by \be \Omega \,=\,
-\frac{1}{2 \pi^2}\int_{0}^{\Lambda} dp\,p^2 \sum_{j=1}^{9}
|\epsilon_j(p)| + \frac{1}{G} (\Delta_1^2+\Delta_2^2+\Delta_3^2)-
\frac{\mu_Q^4}{12 \pi^2}\, ,\label{grandpotential}\ee where
$\Lambda $ is the ultraviolet cutoff, $\epsilon_j(p)$ are the
quasi particle dispersion laws and $G$ is the Nambu-Jona Lasinio
coupling constant.

In  any realistic astrophysical application of these models one
has to impose electric neutrality (the global electric charge must
vanish) and color neutrality (the color charges must vanish as
well). In absence of these conditions, electric and color forces
would destroy quark matter stability. In order to enforce
electrical and color neutrality one has to minimize the grand
potential with respect to $\mu_Q, \mu_3$ and $\mu_8$. Including
the stationary conditions with respect to the gap parameters
$\Delta_1, \Delta_2, \Delta_3$ one ends up with a system of six
equations which must be solved simultaneously. Once this system of
equations is solved one may express $\Delta_1, \Delta_2, \Delta_3$
and the chemical potentials $\mu_Q, \mu_3$ and $\mu_8$ as
functions of $m_s^2/\mu_b$. The superconductive state is gapless
because the equation $\epsilon_j(p)=0$ has solutions for finite
$p$; in other terms, differently from the CFL or the 2SC models,
where the dispersion laws have the form (\ref{dl}), in the gCFL
model there are gapless modes (this happens also in the similar
model with two flavors, the gapless 2SC model \cite{shovkovy}).
This result can be understood as follows. The chemical potentials
of the various quarks are in general different, so that in general
one expects $\delta\mu\neq 0$.  It follows that the generic
quasiparticle dispersion law  has the form \be
\epsilon(p)=\pm\delta\mu+\sqrt{(p-\mu)^2+\Delta^2}\, ,\ee which
shows that, for appropriate values of $\delta\mu$ there are
gapless modes.

Before being accepted, the tentative ground state must pass a
stability test. For gluons this means that the Meissner masses,
obtained by the eigenvalues of the polarization tensor, must be
real. In the CFL phase, with $m_s=0$ the Meissner masses are
degenerate with the value \be m^2_{M}= \frac{\mu_b^2
g^2}{\pi^2}\left(-\frac{11}{36} - \frac{2}{27} \ln 2 + \frac 1 2
\right)\ >\ 0 \, . \label{meissnerCFL}\ee However, for a non zero
strange quark in the gCFL phase the masses are in general
different, due to symmetry breaking, and some of them are actually
imaginary \cite{Casalbuoni:2004tb}. This means that the gCFL state
cannot be the true vacuum state, even if its free energy is
smaller than that of the normal state.

\section{Color Superconductivity and compact stars\label{star1}}
Color superconductivity  might be realized in compact stars. This
 follows from the following considerations.
The BCS critical temperature is given by \be
T_c=0.57\Delta_{BCS}\ee and in QCD $\Delta_{BCS}$ is expected to
range between 20 to 100 MeV, see eq. (\ref{gapcfl}). This result
must be evaluated in the context of the thermal history of
pulsars. These compact stars  are formed  after  a supernova
explosion. The temperature at the interior of the supernova is
about $10^{11}$ K, corresponding to $10$ MeV. Then the star cools
very rapidly by neutrino emission with the temperature going down
to $\sim10^9$-$10^{10}$ K in about one day. Neutrino emission is
then the dominating cooling process for $\sim 10^3$ years. When
the star reaches the temperature of $\sim 10^6$ K, it cools down
due to X-ray and photon emission, so that in a few million years
it reaches a surface temperature $\sim 10^5$ K. Therefore for the
largest part of its existence a neutron star has $T<T_c$, which
implies the possibility of color superconductive states if $\mu$
is large enough. Since the star temperature is much smaller than
the typical BCS energy gap ($T_{\rm n.s.}/\Delta_{BCS}\approx
10^{-6}-10^{-7}$) one can assume that the compact star is
effectively  at zero temperature.

We have shown that QCD favors the formation of BCS condensates in
idealized cases, e.g. two or  three massless flavors of quarks.
However in realistic cases the three quarks have different masses
and, as a consequence, different Fermi momenta. Let us estimate
the order of magnitude of the scales involved in the description
of a neutron star with a quark core. We begin with the simple case
of a free gas of three flavor quarks, assuming that the up and
down quarks are massless and the strange quark has mass $m_s$
\cite{Alford:1999pb}.  We also assume that the weak interactions
are in equilibrium. One gets for quark chemical potentials and
Fermi momenta the results\bea\mu_u&=&\mu-\frac 2
3\mu_e,~~~p_F^u=\mu_u\,,\nn\\
\mu_d&=&\mu+\frac 1 3\mu_e,~~~p_F^d=\mu_d\,,\nn\\
\mu_s&=&\mu+\frac 1 3\mu_e,~~~p_F^s=\sqrt{\mu_s^2-M_s^2}\,,\eea
where $\mu$ is average chemical potential \be\mu=\frac 1
3\,(\mu_u+\mu_d+\mu_s)\,\ee and $\mu_e$ the chemical potential of
the electrons. Notice that \be \sum_{i=u,d,s}\mu_i N_i+\mu_e
N_e=\mu N_q -\mu_e Q\,,\ee where \be
N_q=\sum_{i=u,d,s}N_i,~~~Q=\frac 2 3 N_u-\frac 1
3(N_d+N_s)-N_e\,.\ee This result reflects the fact that the $u$
quark has electric charge $+2/3\,e$, the $d$ and $s$
 charge $-1/3\,e$, and the electron $-e$. The chemical
potential for the electrons is fixed by requiring electrical
neutrality. This corresponds to the following condition for the
grand potential $\Omega$  at zero temperature \be
Q=\frac{\de\Omega}{\de\mu_e}=0\,.\ee For each fermionic species,
omitting the volume factor, $\Omega$ is given by\be \Omega= \frac
1{\pi^2}\int_0^{p_F} p^2(E(p)-\mu)dp\,.\ee In our case we get
\be\Omega=\frac 3{\pi^2}\sum_{i=u,d,s}\int_0^{p_F^i}
p^2(E_i(p)-\mu_i)dp+\frac 1{\pi^2}\int_0^{\mu_e}
p^2(p-\mu_e)dp\,,\ee where \be
E_{u,d}(p)=p,~~~E_s(p)=\sqrt{p^2+m_s^2}\,.\ee

An analytical expression  can be obtained by performing an
expansion up to the order $m_s^4/\mu^4$. One gets \be \mu_e\approx
\frac{m_s^2}{4\mu}\ee and \be \Omega\approx -\frac
3{4\pi^2}\mu^4+\frac
3{4\pi^2}m_s^2\mu^2-\frac{7-12\log(m_s/2\mu)}{32\pi^2}m_s^4\,.\ee
The baryon density is obtained as \be \rho_B=-\frac 1
3\frac{\de\Omega}{\de\mu}=\frac{1}{3\pi^2}\sum_{i=u,d,s}
(p_F^i)^3\,.\ee  With the same approximation as before one finds
\be \rho_B\approx \frac{\mu^3}{\pi^2}\left[1-\frac 1
2\left(\frac{m_s}\mu\right)^2\right]\,.\ee We note that densities
in the core are of the order of $10^{15}\,g/cm^3$, corresponding
to a chemical potential of the order of 400 MeV.

Let us  discuss  the range of values $\mu\sim  400$ MeV of the
average chemical potential, with a strange mass of the order
200-300 MeV (this is the  effective density dependent strange
quark mass. With $m_s=300$ MeV one finds $\mu_e=56$ MeV  with
Fermi momenta \be p_F^u=365\,{\rm MeV},~~~p_F^d=418\,{\rm
MeV},~~~p_F^s=290\,{\rm MeV}\,, \ee and a baryon density about 4.4
times the nuclear matter density. With $m_s=200$ MeV the result is
$\mu_e=25$ MeV
 and \be p_F^u=384\,{\rm
MeV},~~~p_F^d=408\,{\rm MeV},~~~p_F^s=357\,{\rm MeV}\,,\ee and a
baryon density about 5.1 times the nuclear matter density. To go
to baryon densities relevant to the central core of the star, i.e.
densities from 6 to 8 times the nuclear matter density, one needs
to go to higher values of $\mu$ and lower values of $m_s$ where
the difference among of the Fermi momenta is lower. This can be
seen  using the approximate expression for $\mu_e$:\be
p_F^u\approx \mu-\frac{m_s^2}{6\mu},~~~ p_F^d\approx
\mu+\frac{m_s^2}{12\mu},~~~p_F^s\approx
\mu-\frac{5m_s^2}{12\mu}\,,\ee with \be p_F^d-p_F^u\approx
p_F^u-p_F^s\approx \frac{m_s^2}{4\mu}\,.\ee In conclusion the
conditions for the formation of color condensates can be indeed
realized in compact star, more exactly in their cores. The precise
exact nature of the condensed phase is however still matter of
debate. In fact, it is quite likely that the CFL is not favored,
due to the non vanishing strange quark mass. Since the gCFL phase
is unstable, one has to search for an alternative, A possibility
is offered by the inhomogeneous color superconductivity that we
now discuss.

\section{Inhomogeneous color superconductivity: LOFF phase with two flavors}
The relevance of the Larkin-Ovchinnikov-Fulde-Ferrell (LOFF) phase
\cite{LOFF2,Alford:2000ze,Casalbuoni:2003wh,Casalbuoni:2004wm,Casalbuoni:2005zp}
 follows from the fact that,
   for appropriate values of $\delta\mu$, pairing with non-vanishing
total momentum: ${\bf p_1}+{\bf p_2}=2{\bf q}\neq 0$ is favored.
It is a well known fact that if fermions of different chemical
potentials interact in an attractive channel, pairing occurs only
if the difference in chemical  potentials $\delta\mu$ does not
exceed the limit $\Delta_{BCS}/\sqrt 2$, called the
Clogston-Chandrasekhar limit. For $\delta\mu$ much larger than the
Clogston-Chandrasekhar limit the favored state is the normal
phase, but in a window of intermediate values of $\delta\mu$
pairing with non-vanishing total momentum is favored, as first
shown in the context of condensed matter in \cite{LOFF2}. The LOFF
phase is characterized by gap parameters not uniform in space
that, in the simplest case, have the form of a single plane wave:
$\Delta({\bf r})=\Delta\exp(2i{\bf q\cdot r})$. Both the case of
two and three flavours have been discussed in the literature. Let
us begin with the case of 2 flavors , i.e. let us assume that the
strange quark is decoupled due to its (rather) large mass. In
nature flavor symmetry is broken not only explicitly by quark mass
terms, but also by weak interactions. Therefore in the
applications of the color superconductivity one has to take into
account this symmetry breaking; for example in compact stars,
considering only two flavors, isospin is broken by
$\delta\mu=\mu_u- \mu_d\neq 0$, due to the process: \be d\ \to\ u\
e^-\  \bar\nu_e\ . \ee At the equilibrium (including the reaction
$ u\ e^-\ \to d  \ \nu_e$) one has $\delta\mu=-\mu_e$.

To simplify the problem I will consider the case of two massless
quarks with chemical potentials $\mu_u$ and $\mu_u$ given by
\begin{equation}
   \mu_u=\mu+\delta\mu,~~~\mu_d=\mu-\delta\mu\,,
\end{equation}where up and down refer
to flavor. The condensate has the form \begin{equation}
\langle\psi^\alpha_i\psi^\beta_j\rangle\propto\epsilon^{\alpha\beta
3}\epsilon_{ij}\,.\end{equation} Let $\Delta_{BCS}$ be the value
of the  homogeneous BCS condensate. For
$\delta\mu<\Delta_{BCS}/\sqrt 2$, the energetically favored state
is the homogeneous one. To show that for
$\delta\mu>\Delta_{BCS}/\sqrt 2$ the Cooper pair might prefer to
have non zero total momentum, we consider a four-fermion
interaction modelled on one-gluon exchange, that is
  \begin{equation}
 {\it
L}=-\frac 3 8 g\bar\psi\gamma^\mu \lambda^a\psi\,\bar\psi\gamma^\mu
\lambda^a\psi\label{eq:291}\,,
\end{equation}where $\lambda^a$ are Gell-Mann
matrices. In the mean field approximation it reduces
to\begin{equation}{\it L}=-\frac 1 2\epsilon_{\alpha\beta
3}\epsilon^{ij}(\psi_i^\alpha\psi_j^\beta\Delta e^{2i{\bf
q}\cdot{\bf r}}\,+\,{\rm c.c.})\,+\,(L\to R)\,,\end{equation}
where we have defined \begin{equation} \Gamma_S\, e^{2i{\bf
q}\cdot{\bf r}}=-\frac 1 2\,\epsilon^{\alpha\beta 3}\epsilon_{ij}
\langle\psi_\alpha^i\psi_\beta^j\rangle\,,\end{equation}  and
\begin{equation} \Delta=g\Gamma_S\,.\end{equation}
The gap equation has the form
\begin{equation}\Delta=i\frac{g\rho}2\Delta\int\frac{d{\bf
v}}{4\pi}\int_{-\delta}^{+\delta}\,d\xi\, \frac{dE}{2\pi}
\frac{1}{(E-\bar\mu+i\epsilon\,{\rm
sign}E)^2-\xi^2-\Delta^2}\,,\end{equation} where $\rho=4\mu^2/\pi^2$
is the  density of states, $\delta$ the ultraviolet cut-off and
$\xi$ the component of the quark momentum parallel to the Fermi
velocity $\bf v$, measured from the Fermi surface. Moreover
\begin{equation}\bar\mu=\delta\mu-{\bf v}\cdot{\bf q}\,.\end{equation}
Performing the integration over the energy we get \begin{equation}
1=\frac{g\rho}2\int\frac{d{\bf
v}}{4\pi}\int_0^{\delta}\frac{d\xi}{\sqrt{\xi^2+\Delta^2}}\,
\theta(\epsilon-|\bar\mu|)\,,\end{equation} where
\begin{equation}\theta(\epsilon-|\bar\mu|)=1-\theta(-\epsilon-\bar\mu)-\theta(-\epsilon+
\bar\mu)\,.\end{equation} It can be shown that there is a first
order transition in $\delta\mu$, between the homogenous state and
the normal state and  a second order transition between the LOFF
state and the normal one. The first order transition occurs near
 $\delta\mu_1\sim \Delta_{BCS}/\sqrt
2$, while the second order phase transition is at the critical
point $\delta\mu_2=0.754\Delta_{BCS}$ (see \cite{LOFF2}). Near
$\delta\mu_2$ one has
\begin{equation}
\Delta_{LOFF}=\sqrt{1.757\,\delta\mu_2(\delta\mu_2-\delta\mu)}=
1.15\,\Delta_{BCS}\sqrt{\frac{\delta\mu_2-\delta\mu}{\Delta_{BCS}}}\,,\end{equation}which
is non vanishing, and the grand potential is given by
\begin{equation}
\Omega_{LOFF}-\Omega_{normal}=-0.439\,\rho(\delta\mu-\delta\mu_2)^2\,,\end{equation}
showing that the LOFF phase is indeed favored.

In the more general case one can assume the ansatz
\begin{equation}
 \Delta({\bf
r})=\Delta\sum_{m=1}^P\,e^{2i{\bf q}_m\cdot{\bf r}}\
\label{LOgeneral}
\end{equation} (the previous case corresponds to $P=1$). Only approximate solutions can be given for the gap
(\ref{LOgeneral}) either near the second order phase transition,
by the Ginzburg-Landau method, or by an effective gap equation
that captures the main aspects of the structure represented by eq.
(\ref{LOgeneral}).
\subsection{Gap equation in the Ginzburg-Landau approximation}In the Ginzburg-Landau
approximation the gap equation for condensate (\ref{LOgeneral}) is
as follows
\begin{equation}
    \frac{\partial \Omega}{\partial \Delta}=0
\end{equation}with \begin{equation}
\frac{\Omega}{\rho}=\,P\,\frac\alpha 2\Delta^2+\frac \beta 4
\Delta^4
 +\frac\gamma 6\Delta^6
\end{equation}
and
\begin{eqnarray}\alpha&=&\frac2{g\rho}\,(1-\Pi(q))\,,
\hskip1cm\Pi(q)\equiv\Pi({\bf q},{\bf q})\,,\cr \beta&=& -\frac
2{g\rho}\sum_{k,\ell,m,n=1}^P J({\bf q_k},{\bf q_\ell},{\bf
q_m},{\bf q_n})\delta_{{\bf q_k}-{\bf q_\ell} +{\bf q_m}-{\bf
q_n}}\,,\cr \gamma&=&-\frac 2{g\rho}\sum_{k,\ell,m,j,i,n=1}^P K({\bf
q_k},{\bf q_\ell},{\bf q_m} ,{\bf q_j},{\bf q_i},{\bf
q_n})\delta_{{\bf q_k}-{\bf q_\ell} +{\bf q_m}-{\bf q_j}+{\bf
q_i}-{\bf q_n}}\label{gamma1}\ .
\end{eqnarray}
 Here $\delta({\bf q_k}-{\bf q_n}) $
means the Kronecker delta: $\delta_{n,k} $ and the following
definitions have been adopted:
\begin{equation}\Pi({\bf q_1},{\bf q_2})= \,+\,\frac{ig\rho}{2}\int
\frac{d\bf\hat
w}{4\pi}\int_{-\delta}^{+\delta}d\xi\int_{-\infty}^{+\infty}\frac{dE}{2\pi}
\prod_{i=1}^2\, f_i(E,\delta\mu,\{{\bf q}\})\
,\label{eq:143}\end{equation}\begin{equation}
 J({\bf
q_1},{\bf q_2},{\bf q_3},{\bf q_4})= \,+\,\frac{ig\rho}2 \int
\frac{d\bf\hat w}{4\pi} \int_{-\delta}^{+\delta}
d\xi\int_{-\infty}^{+\infty}\frac{dE}{2\pi} \prod_{i=1}^4\,
f_i(E,\delta\mu,\{{\bf q}\})\
,\label{gei}\end{equation}\begin{equation} K({\bf q_1},{\bf
q_2},{\bf q_3},{\bf q_4},{\bf q_5},{\bf q_6})= \,+\,\frac{ig\rho}2
\int \frac{d\bf\hat w}{4\pi}\int_{-\delta}^{+\delta}
d\xi\int_{-\infty}^{+\infty}\frac{dE}{2\pi}\prod_{i=1}^6
f_i(E,\delta\mu,\{{\bf q}\}).\label{kappa}\end{equation}We have put
${\bf w}\,\equiv\,v_F\,\bf\hat w$ and
 \begin{equation}
f_i(E,\delta\mu,\{{\bf q}\})=\frac{1}{E+i\epsilon\,
 {\rm sign}\,E-\delta\mu+(-1)^i[\xi-2\sum_{k=1}^i(-1)^{k}
\bf w\cdot{\bf q_k}]}~;
\end{equation}
moreover the condition $\sum_{k=1}^M(-1)^{k}{\bf q_k}=0$ holds, with
$M=2,4,6$ respectively for $\Pi$, $J$ and $K$. The critical value
$\delta\mu_2$ is obtained by the condition that, at
$\delta\mu=\delta\mu_2$, $\alpha$ vanishes.
 This approximation is adequate to deal with most of the
structures that can be obtained summing several plane waves. This
analysis has been performed by Bowers and Rajagopal in
\cite{Alford:2000ze}.  The most interesting case is offered by the
cubic structures. They are formed either by six plane waves
pointing to the six faces of a cube (the so-called
body-centered-cube, bcc) or by eight plane waves pointing to the
vertices of a cube (face-centered-cube, fcc). In the former case
the grand potential turns out to be bounded from below and smaller
than all the other crystalline structures. In the latter case the
grand potential is unbounded from below, because the coefficient
$\gamma$ is negative. This represents a problem, because it shows
that the analysis is incomplete and higher order terms, e.g. $\sim
\Delta^8$, should be included.
\subsection{Effective gap equation} If one is far away from the
second order phase transition, the Ginzburg Landau approach is not
valid. One can use in this case an effective gap equation obtained
by an appropriate average of the original lagrangian. The
smoothing procedure is described in detail in
\cite{Casalbuoni:2004wm} and we can recall here only the main
points. The method employed uses effective quark fields where the
large part of the quark momentum $\mu\bf v$ has been extracted by
a factor $\exp (i\mu{\bf v})$ ($\bf v$ the Fermi velocity).
Therefore in the condensate term of the lagrangian one has a
factor
\begin{equation}\sum_m\exp\,i(\mu_u{\bf v}_u+\mu_d{\bf v}_d+2{\bf
q}_m)\cdot{\bf r}\ .\label{factor}
\end{equation}
Next we multiply the lagrangian by some appropriate function
$g({\bf r})$ and make an average over a crystal cell. In the gap
equation the relevant integration momenta  are small. Therefore
one can assume that the fields are almost constant in the
averaging procedure. Averaging the factor (\ref{factor}) produces
as a result that the two velocities are antiparallel up to terms
of the order of $\delta\mu/\mu$; moreover the term (\ref{factor})
is a substituted by its average
\begin{equation}\Delta_{E}({\bf
v},\ell_0)=\sum_{m=1}^P\Delta_{eff}\left({\bf v\cdot\bf
n_m},\ell_0\right)\label{eq13}\, ,\end{equation}where $\ell_0$ is
the quasiparticle energy and ${\bf n_m}={\bf q_m}/q$. It can be
shown that this procedure is valid for $\Delta$ not too small,
i.e. far away from the second order phase transition. The exact
form of the function $\Delta_{eff}$ depends on the function
$g({\bf r})$ chosen for the average. In particular $g({\bf r})$
can be chosen in such a way that
\begin{eqnarray}
\Delta_{eff}\,=\Delta\theta(E_u)\theta(E_d)\,=\,{\displaystyle\Bigg
\{ }
\begin{array}{cc}\Delta & \textrm{~~for~~}
(\xi,{\bf v}) \in PR\cr&\cr 0 &
\textrm{~~elsewhere~~} \ .\\
\end{array}
\label{GAP8.1}
\end{eqnarray} where for each plane wave of wave number $\bf n_m$
one has \begin{equation} E_{u,d}=\pm\delta\mu\mp q{\bf n_m} \cdot
{\bf v}+\sqrt{\xi^2+\Delta^2} \label{dispersionFF}
\end{equation} and the pairing region ($PR$) is defined by the condition
$E_u>0,\,E_d>0$. This procedure simplifies the gap equation that
assumes the form
\begin{equation} P \Delta=i\frac{g\rho}{2}\int\frac{d{\bf
v}}{4\pi}\int \frac{d^2\ell}{2\pi} \, \frac{\Delta_{E}({\bf
v},\ell_0)}{\ell_0^2-\ell_\parallel^2-\Delta_{E}^2({\bf
v},\ell_0)}\label{29} \ .\end{equation}
 The energy integration is performed by the residue theorem and
 the phase space is divided into different regions according to
 the pole positions, defined by \begin{equation}
\epsilon=\sqrt{\xi^2+\Delta^2_{E}({\bf
 v},\epsilon)}\ .
\end{equation} Therefore we get\begin{eqnarray}
 P
\Delta\ln\frac{2\delta}{\Delta_0}& =& \sum_{k=1}^P \int\int_{P_k}
\frac{d{\bf v}}{4\pi} d\xi \, \frac{\Delta_{E}({\bf
v},\epsilon)}{\sqrt{\xi^2+\Delta^2_{E}({\bf v},\epsilon)}}\cr
&=&\sum_{k=1}^P\int\int_{P_k}\frac{d{\bf v}}{4\pi} d\xi \,
\frac{k\Delta}{\sqrt{\xi^2+k^2\Delta^2}}\label{290} \
\end{eqnarray}
  where the regions $P_k$ are defined as follows \begin{equation}
P_k=\{({\bf v},\xi)\,|\,\Delta_E({\bf v},\epsilon)=k\Delta\}\
\end{equation} and
 we have made use of the equation\begin{equation}\frac 2{g\rho} =
\ln\frac{2\delta}{\Delta_{BCS}}\end{equation} relating the BCS gap
$\Delta_{BCS}$ to the four fermion coupling $g$ and the density of
states.  The first term in the sum, corresponding to the region
$P_1$, has $P$ equal contributions with a dispersion rule equal to
the Fulde-Ferrel $P=1$ case. This can be interpreted as a
contribution from $P$ non interacting plane waves. In the other
regions the different plane waves have an overlap. The free energy
$\Omega$ is obtained by integrating in $\Delta$ the gap equation.
 At fixed $\delta\mu$, $\Omega
$ is a function of $\Delta$ and $q$, therefore the energetically
favored state satisfies the conditions
\begin{equation} \frac{\partial \Omega}{\partial \Delta} \, =\, 0 \, ,\hspace{1.cm}
\frac{\partial \Omega}{\partial q} \, =\, 0 \,,
\end{equation}and must be the
absolute minimum.   The result of this anlysis is that the
body-centered-cube (bcc) is the favored structure up to
$\delta\mu\approx .95 \Delta_{BCS}$. For larger values of
$\delta\mu<1.32\Delta_{BCS}$ the favored structure is the
face-centered-cube (fcc).

\section{LOFF Phase of QCD with three flavors in the Ginzburg Landau approximation\label{sec:6}}
We shall describe in this section some results for the the LOFF
phase of QCD with three flavours, obtained in the Ginzburg Landau
(GL) approximation \cite{Casalbuoni:2005zp}. We shall limit our
presentation to the case of space modulation given by a single
plane wave. The free energy per unit volume $\Omega$ in the GL
limit is
\begin{equation}
\Omega =\Omega_n+ \sum_{I=1}^3\left(\frac{\alpha_I}{2}\,\Delta_I^2
~+~ \frac{\beta_I}{4}\,\Delta_I^4 ~+~ \sum_{J\neq
I}\frac{\beta_{IJ}}{4}\,\Delta_I^2\Delta_J^2 \right) ~+~ O(\Delta^6)
\label{eq:OmegaDelta11}
\end{equation}
where $\Omega_n$ refers to the normal phase \begin{equation}
 \Omega_n = -\frac{3}{12\pi^2}\left(\mu_u^4+\mu_d^4+\mu_s^4\right) -
\frac{\mu_e^4}{12\pi^2}~, \label{eq:OmegaNorm11222}
\end{equation}
and one has assumed for the condensate the pairing ansatz
\begin{equation}
<\psi_{i\alpha}\,C\,\gamma_5\,\psi_{\beta j}> \propto
\sum_{I=1}^{3}\,\Delta_I({\bf r})\,\epsilon^{\alpha\beta
I}\,\epsilon_{ijI}~\label{cond}
\end{equation}with \be \Delta_I ({\bf r}) = \Delta_I
\exp\left(2i{\bf q_I}\cdot{\bf r}\right)~. \label{eq:1Ws}\ee  In
(\ref{eq:OmegaNorm11222}) $\mu_j$ are chemical potentials for
quarks or for the electron while the coefficients of
(\ref{eq:OmegaDelta11}) can be found in \cite{Casalbuoni:2005zp}.
One works in the approximation of vanishing color chemical
potentials $\mu_3=\mu_8=0$, which is valid for the present case
since they vanish in the normal phase and we work near the secon
order phase transition. $\beta-$equilibrium is imposed together
with the electric neutrality condition
\begin{equation}
-\frac{\partial\Omega}{\partial\mu_e}=0~. \label{eq:electrNeutra111}
\end{equation}These conditions, together with the gap equations, give,
for each value of the strange quark mass, the electron chemical
potential $\mu_e$ and the gap parameters $\Delta_I$. Moreover one
determines ${\bf q_I}$ by solving, together with the gap equation
and Eq. (\ref{eq:electrNeutra111}), also:
\begin{equation}
0~=\ \frac{\partial\Omega}{\partial q_I}=
\Delta_I\frac{\partial\alpha_I}{\partial q_I} +
\Delta_I\sum_{J=1}^{3}\Delta_J^2\frac{\partial\beta_{IJ}}{\partial
q_I}\ ,~~~~~I=1,2,3~. \label{eq:q}
\end{equation}
The condition  (\ref{eq:electrNeutra111}) gives
\begin{equation}
\mu_e \approx \frac{m_s^2}{4\mu}~,\label{mue}
\end{equation} which is
valid up to terms of the order $(1/\mu)$. This result is identical
to the free fermion case, which was expected since one works near
the transition point between the LOFF and the normal phase. It
follows that \be\delta\mu_{du} = \delta\mu_{us}
\equiv\delta\mu\label{ddu}\ee and
\be\delta\mu_{ds}=2\delta\mu~.\label{dds}\ee To evaluate
(\ref{eq:q}), it is sufficient to work at the ${\cal O}(\Delta^2)$,
which leads to  $q = 1.1997|\delta\mu|$.

As to the orientation of $\bf q_j$, the results obtained in Ref.
\cite{Casalbuoni:2005zp} indicate that the favored solution has
$\Delta_1=0$ and therefore ${\bf q_1}=0$. Furthermore ${\bf
q_2}={\bf q_3}$ and $\Delta_2=\Delta_3$. These results are
consequences of the GL limit. In fact, as shown by Eqns. (\ref{ddu})
and (\ref{dds}), the surface separation of $d$ and $s$ quarks is
larger, which implies that $\Delta_1$ pairing is disfavored.  On the
other hand the surface separations of $d$-$u$ and $u$-$s$ quarks are
equal, which implies that $\Delta_2$ and $\Delta_3$ must be almost
equal. Finally $\bf q_2$, $\bf q_3$ are parallel because the pairing
region on the $u-$quark surface at the GL point is formed by two
distinct rings (in the northern and southern emisphere
respectively), while for antiparallel $\bf q_2$, $\bf q_3$ the two
rings overlap, which reduces the phase space available for pairing.

The results one obtains can be summarized as follows. Leaving the
strange quark mass as a parameter (though a complete calculation
would need its determination, considering also the possibility of
condensation in the $q\bar q$ channel) the free energy can be
evaluated as a function of the variable $m_s^2/\mu$. One finds an
enlargement of the range where color superconductivity is allowed,
with the LOFF state having free energy lower than the normal and
the gCFL phases. For example, for $\mu=500$ MeV,  and  with the
coupling fixed by the value $\Delta_0=25$ MeV of the homogeneous
gap for two flavors,  at about $m_s^2/\mu=150$ MeV the LOFF phase
has a free energy lower than the normal one. This point
corresponds to  a second order transition. Then the LOFF state is
energetically favored till the point where it meets the gCFL line
at about $m_s^2/\mu= 128$ MeV. This is a first order transition
since  the gaps are different in the two phases.
\section{Stability of the LOFF Phase of QCD with three
flavors\label{sec:7}}
 In \cite{Ciminale:2006sm} the gluon
Meissner masses in the three flavor LOFF phase of QCD where
computed using the an expansion in $1/\mu$ and the GL
approximation.  At the order of $g^2$ there is a contribution
$g^2\mu^2/(2\pi^2)$ identical for all the eight gluons; this
result for the LOFF phase is identical to those of the normal or
the CFL case as this term is independent of the gap parameters,
see (\ref{meissnerCFL}). Another contribution to the polarization
tensor is given by: \be i\Pi_{ab}^{\mu\nu}(x,y)=\,-\,Tr[\,i\,
S(x,y)\,i\, H_{a}^{\mu}\,i\, S(y,x)\,i\,
H_{b}^{\nu}]\label{eq:Pol}\ee where the trace is over all the
internal indexes; $S(x,y)$ is the quark propagator, and
$H_{a}^{\mu}$ is the vertex. The quark propagator $S$ has
components $S^{ij}$ ($i,\,j=1,2$), where each $S^{ij}$ is a
$9\times9$ matrix in the color-flavor space. At the fourth order
in $\Delta$ one has: \be
S^{11}=S_{0}^{11}+S_{0}^{11}\,{\Delta}\,\left[S_{0}^{22}{\Delta^{\star}}\,
\left(S_{0}^{11}+
S_{0}^{11}\,{\Delta}\,S_{0}^{22}\,{\Delta^{\star}}\,
S_{0}^{11}\right)\right]\,,\label{s11}\\
S^{21}=S_{0}^{22}\,{\Delta^{\star}}\,\left(S_{0}^{11}+S_{0}^{11}{\Delta
}\,S_{0}^{22}{\Delta^{\star}}\,S_{0}^{11} \right)\,,\ee where
$S_0^{ij}$ is the $18\times18$ matrix:\begin{equation}
S_0=\left(\begin{array}{cc}
              [S_0^{11}]_{AB} & 0 \\
              0 & [S_0^{22}]_{AB}
            \end{array}
\right)=\delta_{AB}\left(\begin{array}{cc}
              \left(p_0-\xi+\bar\mu_A \right)^{-1} & 0 \\
              0 & \left(p_0+\xi-\bar\mu_A \right)^{-1}
            \end{array}
\right),\label{eq:FreeProp}
\end{equation}
$A,\,B=1,\cdots 9$ are indexes in the basis ${A} \ = \ \left({u}_r,
{d}_g, {s}_b, {d}_r, {u}_g, {s}_r, {u}_b, {s}_g, {d}_b\right)$,
${\Delta}$ and ${\Delta^\star}$ are $9\times9$ matrices containing
the gap parameters $\Delta_I$, $p_0$ is the energy, $\xi=|{\bf
p}|-\mu$,
 $\bar\mu_{A} \ =\mu_A-\mu$.
 $S^{12}$ and $S^{22}$ are obtained by the changes $11\leftrightarrow 22$ and
 ${\Delta}\leftrightarrow{\Delta^\star}$.

  In the region where the use of the GL expansion is
justified, i.e.  $128$ MeV $<m_s^2/\mu < 150$ MeV~ all the squared
gluon Meissner masses were found positive and therefore the LOFF
phase of three flavor QCD is free from the chromo-magnetic
instability.
\section{Neutrino emission by pulsars and the LOFF state}
As noted above neutrino emission due to direct Urca processes,
when kinematically allowed, is the most important cooling
mechanism for a young neutron star. When the temperature of the
compact star is of the order of $\sim 10^{11}$K neutrinos are able
to escape, which produces a decrease in temperature. For smaller
temperatures, e.g. below $10^9$K, the direct nuclear Urca
processes $n \to p + e + \bar\nu_e$ and $e^- +p \to n + \nu_e$,
would produce rapid cooling. However they are not kinematically
allowed, because energy and momentum cannot be simultaneously
conserved. Therefore only modified Urca processes, involving
another spectator particle, can take place. The resulting cooling
is less rapid because neutrino emission rates turn out to be
$\varepsilon_{\nu} \sim T^8$, much smaller than the emission rate
$\varepsilon_{\nu} \sim T^6$ due to direct Urca processes.

These considerations apply to stars containing only nuclear
matter. If hadronic densities in the core of neutron stars are
sufficiently large, as we have seen, the central region of the
star may comprise  deconfined quark matter in a color condensed
state. Therefore direct Urca processes involving quarks, i.e. the
processes $d \to u + e^- +\bar\nu_e$ and $u + e^- \to d +\nu_e$,
may take place and contribute to the cooling of the star. In the
color-flavor locked phase, in which light quarks of any color
 form Cooper pairs with zero total
momentum, and all fermionic excitations are gapped, the
corresponding neutrino emissivity and the specific heat $C$ are
suppressed by a factor $e^{-\Delta/T}$ and the cooling is
distinctly less rapid compared to quarks in the normal phase.
However at densities relevant for compact stars the quark number
chemical potential $\mu$ should be of the order of  500 MeV and
effects due to  the strange quark mass $m_s$ are relevant. In
\cite{anglani} a calculation  of neutrino emission rate and the
specific heat of quark matter in the LOFF superconductive phase
with three flavors was performed. It was shown that, due to the
existence of gapless modes in the LOFF phase, a neutron star with
a quark LOFF core cools faster than a star made by nuclear matter
only. This follows from the fact that in the LOFF phase neutrino
emissivity and quark specific heat are parametrically similar to
the case of unpaired quark matter ($\varepsilon_{\nu} \sim T^6$
and $C\sim T $ respectively). Therefore the cooling is similar to
that of a star comprising unpaired quark matter. These results are
still preliminary since  the simple ansatz of a single plane wave
 should be substituted by a more complex behavior and it remains
 to be seen if a more complete calculation would confirm the
 results of \cite{anglani}.

\acknowledgments

 I would like to thank R. Anglani, R. Casalbuoni,
M. Ciminale, R. Gatto, N. Ippolito, M. Mannarelli and M. Ruggieri
for a most fruitful collaboration on the themes covered by this
review.


\begin{thebibliography}{0}
\bibitem{others}\BY{Bailin D.\atque Love
A.}\IN{Phys. Rept.}{107}{1984}{325}.
\bibitem{rassegne}\BY{Rajagopal K. \atque
Wilczek F.} in \TITLE{Handbook of QCD} edited by \NAME{Shifman
M.,} {World Scientific} (2001),  pp. 2061
[arXiv:hep-ph/0011333];~\BY{Nardulli G.}~\IN{Riv. Nuovo
Cimento}{25N3}{2002}{1}{~[arXiv:hep-ph/0202037]}; \BY{Schafer~T.}
    arXiv:hep-ph/0304281.

\bibitem{NJL}\BY{Nambu Y.  \atque  Jona Lasinio G.}\IN{Phys. Rev.}
{122}{1961}{345}; \SAME{124}{1961}{246}.
%

\bibitem{alford}\BY{Alford M., Rajagopal K. \atque Wilczek F.}
\IN{Phys. Lett. B}{422}{1998}{247}{~ [arXiv:hep-ph/9711395]}.
%
\bibitem{wilczekcfl}
\BY{Alford M., Rajagopal K. \atque Wilczek F.}\IN{Nucl. Phys.
B}{537}{1999}{443}{~[arXiv:hep-ph/9804403]}.

\bibitem{Alford:2003fq}
\BY{Alford M., Kouvaris C. \atque Rajagopal K.} \IN{Phys. Rev.
Lett.}{92}{2004}{222001}{~[arXiv:hep-ph/0311286]}; \IN{Phys. Rev.
D}{71}{2005}{054009}{~[arXiv:hep-ph/0406137]}.

\bibitem{Casalbuoni:2004tb}
\BY{Casalbuoni R.,~Gatto  R.,~Mannarelli  M.,~Nardulli  G. \atque
Ruggieri~M.}\IN{Phys. Lett. B}{605}{2005}{362};~
\IN{Erratum-ibid.}{615}{2005}{297}{~[arXiv:hep-ph/0410401]}; \BY
{Fukushima~K.}\IN{Phys. Rev.
D}{72}{2005}{074002}{~[arXiv:hep-ph/0506080]}
\bibitem{LOFF2}\BY{Larkin, A. I. \atque
Ovchinnikov,Yu. N.}\IN{Zh. Eksp. Teor.
Fiz.}{47}{1964}{1136}~(\IN{Sov. Phys.
JETP}{20}{1965}{762});~\BY{Fulde, P. \atque  Ferrell, R.
A.}\IN{Phys. Rev. A}{135 }{1964}{550}.
\bibitem{Alford:2000ze}
\BY{Alford M.,~Bowers J.~A \atque Rajagopal  K.}\IN{Phys. Rev. D
}{63}{2001}{074016}{~[arXiv:hep-ph/0008208]}; \BY{Bowers J.~A.
\atque Rajagopal K.}\IN{Phys.
Rev.D}{66}{2002}{065002}{~[arXiv:hep-ph/0204079]}.
\bibitem{Casalbuoni:2003wh}
\BY{Casalbuoni R. \atque Nardulli G.}\IN{
 Rev. Mod. Phys.}{76}{2004}{263}{~[arXiv:hep-ph/0305069]}.
 \bibitem{Casalbuoni:2004wm}\BY{Casalbuoni  R.,~Ciminale M.,~Mannarelli M.,~Nardulli G.,~Ruggieri M. \atque
 R.~Gatto}
  \IN{Phys.\ Rev. D}{ 70}{2004}{054004}{
  ~[arXiv:hep-ph/0404090]}

  \bibitem{Casalbuoni:2005zp}\BY{Casalbuoni R.,~Gatto R.,~Ippolito N.,~Nardulli G. \atque~Ruggieri
 M.}\IN{Phys. Lett. B}{627}{2005}{89}~(\IN{Erratum-ibid.}{634}{2006}{565}
 {~[arXiv:hep-ph/0507247]};~\BY{Mannarelli M.,~Rajagopal  K. \atque~Sharma
  R.}\IN{Phys. Rev. D}{73}{2006}{114012}~[{arXiv:hep-ph/0603076]};~\BY{Rajagopal K.~
  \atque~Sharma R.}{~[arXiv:hep-ph/0605316]}.
\bibitem{Rapp}\BY{Rapp R.,~Schafer T.,~Shuryak E.~V. \atque Velkovsky
M.}\IN{Phys.\ Rev.\ Lett.}{81}{1998}{53}{ [arXiv:hep-ph/9711396]}.
\bibitem{shovkovy}\BY{Shovkovy I. \atque ~Huang M.}\IN{Phys.\ Lett.
B}{564}{2003}{205}{ [arXiv:hep-ph/0302142]}; \IN{Phys.\ Rev.
D}{70}{2004}{051501}{ [arXiv:hep-ph/0407049]};\IN{Phys.\ Rev.
D}{70}{2004}{094030}{ [arXiv:hep-ph/0408268]}.
  \bibitem{Rajagopal:2000rs} \BY{Rajagopal K. \atque and Shuster E.}
  \IN{Phys. Rev. D}{62}{2000}{085007}{\ [arXiv:hep-ph/0004074]}.

\bibitem{Alford:1999pb} \BY{Alford M. G., Berges J.\atque Rajagopal K.}
  \IN{Nucl. Phys. B}{571}{2000}{269}.

  \bibitem{Ciminale:2006sm}
\BY{Ciminale~M.,~Nardulli G.,~Ruggieri M. \atque~Gatto  R.}
\IN{Phys. Lett. B}{636}{2006}{317}{
  ~[arXiv:hep-ph/0602180]}.

\bibitem{anglani}
\BY{Anglani~R.,~Mannarelli M., Nardulli G.~ \atque~Ruggieri M.}
  ~[arXiv:hep-ph/0607341].
\end{thebibliography}
\end{document}